\documentstyle[prc,preprint,tighten,aps]{revtex}
\begin{document}
\draft
\title{
The \bbox{^7}Be$(p,\gamma)$\bbox{^8}B cross section and the
properties of \bbox{^7}Be}
\author{A.\ Cs\'ot\'o, K.\ Langanke, S.\ E.\ Koonin,
and T.\ D.\ Shoppa}
\address{W.~K. Kellogg Radiation Laboratory, 106-38, California
Institute of Technology, Pasadena, California 91125}
\date{\today}

\maketitle

\begin{abstract}
\noindent
We study the nonresonant part of the $^7$Be($p,\gamma$)$^8$B
reaction using a three-cluster resonating group
model that is variationally converged and virtually
complete in the $^4$He+$^3$He+$p$ model space. The importance
of using adequate nucleon-nucleon interaction is demonstrated.
We find that the low-energy astrophysical $S$-factor
is linearly correlated with the quadrupole moment of $^7$Be.
A range of parameters is found where the most important
$^7$Be and $^7$Li properties are reproduced simultaneously;
the corresponding $S$-factor at $E_{\rm cm}=20$ keV is
$24.6-26.1$ eV$\cdot$b.
\end{abstract}

\narrowtext

The flux of high-energy neutrinos generated in the solar core
is directly proportional to the $^7$Be($p,\gamma$)$^8$B
reaction rate. Thus, knowledge of $S_{17}$, the
$^7$Be($p,\gamma$)$^8$B $S$-factor at solar energies
(center-of-mass energy $E\approx 20$ keV), is crucial to
conclusions drawn from present (Homestake, Kamiokande) and
future (SNO, Super\-kamiokande) solar neutrino experiments
\cite{Bahcall,Hata}. Despite extensive experimental efforts,
the $^7$Be($p,\gamma$)$^8$B cross section is still the most
uncertain nuclear input to the standard solar model
\cite{Bahcall,SSM}, due to a significant spread
among the values of $S_{17}$ deduced from the various
experiments (direct capture \cite{Filippone}: $S_{17}=18-28$
eV$\cdot$b and Coulomb break-up \cite{Motobayashi}:
$S_{17}=16.7\pm3.5$ eV$\cdot$b). Theoretical estimates
also vary ($S_{17}=16-30$ eV$\cdot$b) \cite{Kh1},
making these predictions rather unreliable.

The aim of this Letter is to constrain more tightly the
theoretical value of $S_{17}$. To this end, we study the
$^7$Be($p,\gamma$)$^8$B reaction in a microscopic
three-cluster ($^4$He+$^3$He+$p$) approach. This model is
currently the closest approximation to a full solution
of the microscopic eight-nucleon problem with a consistent
treatment of bound and scattering states. As we will
demonstrate below, our approach is superior to all
previous studies of the low-energy $^7$Be($p,\gamma$)$^8$B
reaction.

Adopting a microscopic three-cluster ($^4$He+$^3$He+$p$) ansatz
for the eight-nucleon system, our trial function reads
\begin{equation}
\Psi =\sum_{(ij)k,S,l_1,l_2,L}
{\cal A}\left \{\left [ \left [\Phi^i(\Phi ^j\Phi^k)
\right ]_S \chi ^{i(jk)}_{[l_1l_2]L}(
\mbox{\boldmath $\rho $}_1,
\mbox{\boldmath $\rho $}_2)\right ]_{JM}\right \},
\label{wfn}
\end{equation}
where the indices $i,j$, and $k$ denote any  of the labels
$^4$He, $^3$He, and $p$. In (\ref{wfn}) ${\cal A}$ is the
intercluster antisymmetrizer, the cluster internal states
$\Phi$ are translationly invariant harmonic oscillator shell
model states, the \mbox{\boldmath $\rho $} vectors are the
intercluster Jacobi coordinates, and [$\dots$] denotes angular
momentum coupling. In the sum over $S,l_1,l_2$, and $L$ we
include all angular momentum configurations of any
significance. This same model was used in \cite{B8} in the
study of the ground state of $^8$B; further details on the
model space and other aspects can be found there. The
intercluster dynamics is determined by inserting (\ref{wfn})
into the eight-nucleon Schr\"odinger equation using the
two-nucleon strong and Coulomb interactions. In addition
to the full model space calculation, which contains all three
possible arrangements of the three clusters, we also present
a restricted calculation involving only ($^4$He,$^3$He)$p$
configurations ($^7$Be+$p$ type model space), analogous to
simple $^7$Be+$p$ potential model studies, e.g.\ \cite{Kim}.

It is well known that the low-energy $^7$Be($p,\gamma$)$^8$B
cross section is strongly dominated by E1 capture. (Previous
microscopic calculations have shown that M1 capture only plays
a role in the vicinity of the $1^+$ resonance at $E=640$ keV
and is negligible at astrophysical energies \cite{Baye}, while
E2 capture is tiny at $E<500$ keV and can safely be ignored.)
We have therefore calculated the E1 capture cross section into
the $^8$B ground state in perturbation theory (as outlined for
example in \cite{Baye}), describing the initial scattering
states and the $^8$B ground state by the many-body wave
functions determined in our microscopic three-cluster approach.

The capture cross section contains the bound $^8$B and
the scattering $^7$Be+$p$ wave functions. At low energies,
deep below the Coulomb barrier, the capture takes place at
large $^7$Be-$p$ distances, which means that these wave
functions must be accurate to distances of a few hundred
fermis, which requires a reliable method to
determine the unknown relative motion functions $\chi$
in (\ref{wfn}). We expand these functions in terms of
products of basis functions of the Jacobi coordinates,
which allow us to reduce the three-cluster wave functions
(\ref{wfn}) to equivalent two-cluster forms \cite{Tang}.
For example, if we have $N$ basis functions between the
$^4$He and $^3$He clusters in the ($^4$He,$^3$He)$p$
partition, after this reformulation we get a $^7$Be+$p$
two-cluster wave function with the $^7$Be ground state and
$N-1$ (unphysical, continuum) excited states.

We use the variational Siegert method to determine the $^8$B
boundstate \cite{Gir}. The trial state contains tempered Gaussian
functions \cite{He6} plus a term with the correct outgoing
Whittaker assymptotics, $W_{\eta,l}^{(+)}(k\rho)$, in the
$^7$Be+$p$ partitions. (Here $k$ is the wave number
corresponding to the $^8$B binding energy relative to the
$^7$Be+$p$ threshold.) Using such a trial function in a linear
variational method leads to a transcendental equation for the
binding energy, which can be solved iteratively. To be able to
calculate every many-body matrix element analytically, we match
the external Whittaker functions with internal Gaussians, using a
modified version of the technique described in \cite{Kamimura}.
The accuracy of this procedure is better than 1-2\% in $S_{17}$.
For comparison, the uncertainity coming from the different
possible ways of handling the proton-neutron mass difference
is estimated to be 2.5\% \cite{Kajino}.

The scattering wave functions were calculated using the
variational Kohn-Hulth\'en method \cite{Kamimura}, which
ensures the correct scattering assymptotics. To achieve
high accuracy we avoid the use of complex wave functions
and so neglect channel coupling between different
angular momentum channels. Note that this approximation
is certainly justified at astrophysical energies where the
capture occurs far outside the range of the strong forces.
The present scattering solution is numerically
well-conditioned for $E>3$ keV, and its accuracy is better
than 0.1 \%. The technical details and further physical
implications of the model will be published elsewhere
\cite{Csotofut}.

The bulk of our calculations use the Minnesota (MN) effective
nucleon-nucleon interaction \cite{Minnesota}, which contains
central and spin-orbit terms. This force reproduces the most
important properties of the low-energy $N+N$ and $^4$He+$N$
scattering phase shifts and, as we show below, appears to be
best suited for the problem at hand. However, we will also
compare these results to those obtained with other effective
NN interactions. Note that the tensor component of the effective
NN interaction in microscopic cluster models is not well
constrained \cite{B8} and is usually ignored. Nevertheless
we have also performed a calculation including a tensor
force, which, at the least, gives the correct low-energy order
of the triplet-odd $N+N$ phase shifts \cite{B8}. As is customary,
we will present our results in terms of the astrophysical
$S$-factor
\begin{equation}
S(E)=\sigma(E)E\exp\Big [2\pi\eta(E)\Big ],
\end{equation}
where $\eta$ is the Sommerfeld parameter.

The free parameters in our model are the size parameter
($\beta$) in the $^4$He and $^3$He cluster model functions
(technical reasons force us to use the same value for both
$^4$He and $^3$He), the exchange mixture parameter of the
central part of the effective NN interaction, and the strength
of the spin-orbit force. It is generally preferable to adjust
these parameters to independent data. However, a meaningful
study of the $^7$Be($p,\gamma$)$^8$B reaction at low energies
requires the {\em exact} reproduction of the experimental $^8$B
binding energy (137 keV), which we have guaranteed by the
appropriate choice of the exchange mixture parameter. The
strength of the spin-orbit force was adjusted to the
experimental splitting between the $3/2^-$ and $1/2^-$
$^7$Be states. We have varied $\beta$, thus changing our
description of the $^7$Be properties.

As is demonstrated by the open circles in Fig.\ \ref{fig1},
$S_{17}$ scales linearly with the quadrupole moment of $^7$Be,
$Q_{^7\rm Be}$. This linear dependence can be understood as
follows. As the capture process takes place at very large
$^7$Be-$p$ distances, where the bound state wave function is
proportional to the fixed Whittaker function,
$\Psi_{^8{\rm B}}(\rho)\cdot\rho=\bar cW_{\eta,l}^{(+)}(k\rho)$,
the low-energy cross section depends almost exclusively on the
square of the asymptotic normalization factor $\bar c$. Let us
compare calculations with different $^7$Be wave functions,
which give different $^7$Be radius, quadrupole moment, etc.,
but with fixed binding energy of $^8$B. The effective local
potentials between $^7$Be and $p$ have different radii, which
means that the height of the Coulomb barrier is larger if the
potential radius (and the $^7$Be radius) is smaller. Consequently,
the probability of finding the proton in the outside region
decreases as the size of the $^7$Be nucleus becomes smaller.
But as the shape of the external wave function is fixed, this
smaller probability must stem from a smaller normalization
constant $\bar c$. It is easy to see that this leads $\bar c^2$,
and consequently $S_{17}$, to be linearly
proportional to either $r^2_{^7\rm Be}$ or $Q_{^7\rm Be}$.
Note that this relation is not changed if a tensor
component is added to the MN interaction (see triangle
in Fig.\ \ref{fig1}). We find the same linear
$S_{17}$-$Q_{^7\rm Be}$ relation in our truncated
calculation considering only the $^7$Be+$p$ model space.
Results of these restricted calculations are shown in
Fig.\ \ref{fig1} as full circles.

Unfortunately the linear relation is not sufficient to determine
$S_{17}$ indirectly by measuring the $^7$Be quadrupole moment,
as this relation depends upon the effective NN interaction used.
To demonstrate this we have performed calculations within the
$^7$Be+$p$ model space using the Volkov force V2 and the modified
Hasegawa-Nagata (MHN) force, both of which have been used in previous
microscopic cluster calculations of the $^7$Be($p,\gamma$)$^8$B
reaction at low energies \cite{Baye,Baye2,Johnson}. While both
forces also show the linear dependence between $S_{17}$ and the
$^7$Be quadrupole moment, the V2 force yields larger values for
$S_{17}$ for a given $Q_{^7\rm Be}$ (diamonds in Fig.\
\ref{fig1}), while the MHN force yields smaller values (squares).
These differences can be traced to the different quality of
the description of the $N+N$ systems (phase shifts, energy and
radius of the deuteron) by these forces. For example, while
the MN force well reproduces the experimental deuteron energy
and radius, the V2 force underbinds the deuteron by 1.6 MeV
(however, it unphysically binds the singlet dinucleon states)
and the MHN force overbinds it by 4.4 MeV. We note that
the M3Y interaction, which was used in Ref.\ \cite{Mukh} in
an external capture approach to predict a very small
$^7$Be(p,$\gamma$)$^8$B cross section ($S_{17}=$16.5 eV$\cdot$b),
also overbinds the deuteron. From this discussion we
conclude that the Minnesota force is by far the most carefully
constructed force available in the cluster literature; we
will adopt it in the following. Relatedly, cluster calculations
using the MHN and V2 forces should be regarded with care.

Accepting  the MN force as adequate for the eight-nucleon
problem, our result for $S_{17}$ could be read off Fig.\
\ref{fig1} if the $^7$Be quadrupole moment were known.
Absent this information, we will estimate a best
$S_{17}$ value by constraining the $^4$He and $^3$He cluster
size parameter to reproduce (i) the binding energy of $^7$Be
with respect to $^4$He+$^3$He; (ii) the squared sum of the
$^4$He and $^3$He radii; (iii) the quadrupole moment of
$^7$Li (as a surrogate for the unknown quadrupole moment
of the analog nucleus $^7$Be). These requirements ensure
that both the $^7$Be bound states and the $^4$He-$^3$He
relative motion are well described. The second requirement
is fulfilled by choosing $\beta=0.4$ fm$^{-2}$. With this
choice, the $^7$Be ground state is slightly underbound by
200 keV, while the quadrupole moment of $^7$Li is calculated
as $Q=-4.10$ e$\cdot$fm$^2$, to be compared with the
experimental value $-4.05\pm$0.08 e$\cdot$fm$^2$ \cite{quad}.

We conclude that the three requirements above can be reasonably
fulfilled simultaneously. The corresponding $S_{17}$ value is
then 26.1 eV$\cdot$b, while the $^7$Be quadrupole moment is $-6.9$
e$\cdot$fm$^2$. Our approach then calculates the quadrupole
moment of $^8$B as 7.45 e$\cdot$fm$^2$, while the experimental
value is (6.83$\pm$0.21) e$\cdot$fm$^2$ \cite{Minamisono}. Even
if one concludes from a comparison of our $^7$Li and $^8$B
quadrupole moments with experiment that our $^7$Be quadrupole
moment is also slightly too large, we note that a 10$\%$
reduction in this quantity would only decrease $S_{17}$
to 23.5 eV$\cdot$b.

If we use the same cluster size parameter in the restricted
$^7$Be+$p$ space as in the full calculation ($\beta =0.4$ fm$^2$),
we find that the $^7$Be nucleus is overbound (by
600 keV), while its quadrupole moment is reduced to $Q=-6.0$
e$\cdot$fm$^2$. The quadrupole moments of $^7$Li ($Q=-3.46$
e$\cdot$fm$^2$) and $^8$B ($Q=6.55$ e$\cdot$fm$^2$) are
slightly smaller than the experimental values. In this
restricted calculation we find $S_{17}$ to be 24.6 eV$\cdot$b.

If we consider that both the full and restricted $^7$Be+$p$ model
spaces predict the same linear dependence of $S_{17}$ on the
$^7$Be quadrupole moment and that these calculations bracket
the experimental $^7$Li and $^8$B quadrupole moments, we
conclude that the microscopic three-cluster calculations predict
$S_{17}$ to be in the range between 24.6 and 26.1 eV$\cdot$b.
This does not support speculations that $S_{17}$ might be
noticeably smaller ($S_{17}=$16.5, 16.9, and 17 eV$\cdot$b in
\cite{Mukh}, \cite{Zubarev}, and \cite{Barker}, respectively)
than currently accepted in the standard solar model. We
note, however, that $S_{17}$ deduced from our model is
consistent with the value deduced from the direct capture
data (22$\pm$2 eV$\cdot$b \cite{Johnson}, and
24$\pm$2 eV$\cdot$b \cite{Filippone}, respectively).

Less elaborate microscopic cluster calculation have been
presented in Refs.\ \cite{Baye,Assenbaum,Johnson,Baye2}.
While the two earlier studies \cite{Baye,Assenbaum} were
restricted to a simple $^7$Be+$p$ model space, Ref.\
\cite{Baye2} recently improved these studies by including
a $^5$Li+$^3$He rearrangement channel. However, in
\cite{Baye2} the $^7$Be nucleus is described by only
one Gaussian basis function between $^4$He and $^3$He, which
means that the three-cluster wave function is not free for
the variational method. Letting the trial function more
flexible would result in the collapse of the artificially
fixed wave function. Moreover, in \cite{Baye2} the description
of the $^7$Be nucleus is rather unphysical, as it is unbound
relative to the $^4$He+$^3$He threshold. In \cite{Johnson}
there are two basis functions for $^7$Be, with carefully
chosen parameters, and the most important angular momentum
configurations of the $^7$Be+$p$ type partition are present.
In the present model we use six states for $^7$Be (and ten
in the $^7$Be--$p$ relative motion, and six in all other
relative motions) and include all relevant angular momentum
channel. Our test calculations showed that the present
three-cluster model space is virtually complete (see also
the discussion in \cite{B8}), which means that our results
do not contain the side-effects of an unconverged or
incomplete model. Although the incompleteness of the
previous works makes the comparison difficult, our results
are qualitatively in good agreement with \cite{Baye2} and
\cite{Johnson}.

In Fig.\ \ref{fig2} we show the energy-dependence of the
$S$-factor, calculated with the $^7$Be+$p$ model space, the
MN force, and $\beta=0.4$ fm$^2$. At low energies our
calculated $S$-factor is in rather close agreement with
the direct capture data of Ref.\ \cite{Kav}. Although
our $S$-factor also agrees well with the data of
\cite{Vaughn} for $E>1$ MeV, its energy dependence
might change, if the coupling between the different
angular momentum channels in the $^7$Be+$p$ scattering
is taken into account.

In summary, we have studied the $^7$Be($p,\gamma$)$^8$B
reaction in a microscopic model. At low energies this
model is virtually complete in the three-cluster model
space. We found that the low-energy astrophysical
$S$-factor is strongly correlated with the properties of
$^7$Be (e.g., its quadrupole moment). For a set of
parameters that reproduce simultaneously the most
important properties of $^7$Be, $^7$Li and $^8$B we
predict $Q_{^7{\rm Be}}$ to be between $-6.0$ e$\cdot$fm$^2$
and $-6.9$ e$\cdot$fm$^2$ and find $S_{17}=24.6-26.1$ eV$\cdot$b,
in good agreement with direct capture results and the currently
accepted value in the standard solar model. Our calculation,
thus does not support the recently suggested smaller values
for $S_{17}$ \cite{Mukh,Zubarev}. If it turns out that the
$S$-factor is considerably lower than our present value,
then the present three-cluster approach is inappropriate and
physics beyond our model (larger eight-body model space, improved
effective interaction) has to be invoked. We have also shown
that the NN interaction used in cluster models must be
carefully chosen. Although we found that only the Minnesota
force was suitable for the present work, the construction and
use of other high quality interactions would be indespensable.
We also note that a precise measurement of the $^7$Be quadrupole
moment or radius could test the self-consistency of our
conclusions.

This work was supported by the Fulbright Foundation (A.\ C.) and NSF
Grant Nos. PHY90-13248 and PHY91-15574 (USA), and by OTKA Grant Nos.\
3010 and F4348 (Hungary). We thank Dr.\ E.\ Kolbe for useful
discussions and Dr.\ Z.\ Papp for his generous help in numerical
methods.

\begin{figure}
\caption{The astrophysical $S$-factor of the
$^7$Be($p,\gamma$)$^8$B reaction as a function of the
negative of the $^7$Be quadrupole moment. The symbols
are explained in the text.}
\label{fig1}
\end{figure}

\begin{figure}
\caption{Energy dependence of the $^7$Be($p,\gamma$)$^8$B
astrophysical $S$-factor. The symbols denote the experimental
data of Ref.\ \protect\cite{Kav} (open circles), Ref.\
\protect\cite{Brad} (filled circles), and Ref.\
\protect\cite{Vaughn} (squares). The inset shows the
low-energy part on a magnified scale. See the error bars in the
original references.}
\label{fig2}
\end{figure}
\end{document}